# Efficient utilization of Channels using Dynamic Guard Channel Allocation with Channel Borrowing Strategy in Handoffs

Alagu S[1], Meyyappan T[2]

[1]Research Scholar, Department of Computer Science and Engineering
Alagappa University, Karaikudi, Tamilnadu, India
sivaalagu@hotmail.com

[2]Associate Professor, Department of Computer Science and Engineering
Alagappa University, Karaikudi, TamilNadu, India
meyslotus@yahoo.com

## ABSTRACT

*User mobility in wireless data networks is increasing because of technological advances and the desire for voice and multimedia applications. These applications, however, require fast handoffs between base stations to maintain the quality of the connections. In this paper, the authors describe the use of novel and efficient data structure which dynamically allocates guard channel for handoffs and introduces the concept of channel borrowing strategy. The proposed scheme allocates the guard channels for handoff requests dynamically, based on the traffic load for certain time period. A new originating call in the cell coverage area also uses these guard channels if they are unused. Our basic idea is to allow Guard channels to be shared between new calls and handoff calls. This approach maximizes the channel utilization. The simulation results prove that the channel borrowing scheme improves the overall throughput.*

## KEYWORDS

*Guard Channel, Dynamic, Channel Borrowing, Handoff, QoS*

## 1. INTRODUCTION

Mobility is the most important feature of wireless cellular communication system. Usually continuous service is achieved by supporting handoff (or handover) from one cell to another. Handoff is the process of changing the channel (frequency, timeslot, spreading code, or combination of them) associated with the current connection while a call is in progress [3]. It is often initiated either by crossing a cell boundary or by deterioration in quality of signal in the current channel. Poorly designed handoff schemes tend to generate very heavy signaling traffic and thereby a dramatic decrease in the quality of service (QoS).

### 1.1. Mobile Communication system

A cellular network allows cellular subscribers to wander anywhere in the region and remain connected to the Public Switched Telephone Networks (PSTN) via their wireless mobile devices. A cellular network has a hierarchical structure and it is formed by connecting Mobile Stations (MS), Base Station (BS) and Mobile Switching Centre (MSC). The Base Station serves a cell which could be few kilometers in diameter. The cell is a part of a larger region, which has been partitioned into smaller regions such that there is a base station serving each cell. All BSs within the cluster are connected to MSC. Each MSC of a cluster is then connected to the MSC of other clusters and then to PSTN. The architecture of the same is depicted in figure 1. The

MSC stores information about the subscribers located within the cluster and is responsible for directing calls to them [5].

Neighboring cells overlap with each other to ensure the continuity of communications when the users move from one cell to another. Certain number of channels is allocated to each base station. A channel in the system can be thought of as a fixed frequency bandwidth (FDMA), a specific time-slot within a frame (TDMA), or a particular code (CDMA), depending on the multiple access technique used. BSs and MSCs take the responsibility of allocating channel resources to mobile stations [6].

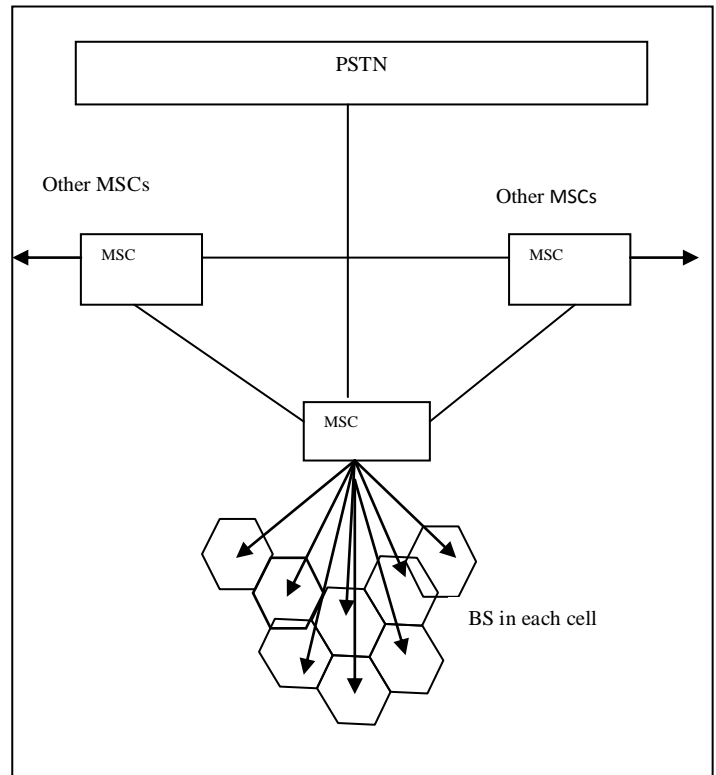

Figure 1. Mobile Communication System

## 1.2. Handoff

Handoff (also called Handover) is the mechanism that transfers an ongoing call from one cell to another as a user moves through the coverage area of a cellular system. The handoff process is initiated by the issuing of handoff request. The power received by the MS from BS of neighboring cell exceeds the power received from the BS of the current cell by a certain amount. This is a fixed value and is called the handoff threshold. For successful handoff, a channel must be granted to handoff request before the power received by the MS reaches the receiver's threshold. The handoff area is the area where the ratio of received power levels from the current and the target BS's is between the handoff and the receiver threshold [4][7][9]. Each handoff requires network resources to reroute the call to the new base station. Minimizing the expected number of handoff minimizes the switching load. Another concern is delay. If the handoff does not occur quickly, the Quality of Service (QoS) may degrade below an acceptable level. Minimizing delay also minimizes co-channel interference. During handoff there is brief service interruption. As the frequency of these interruptions increases, the perceived QoS is reduced. The chance of dropping a call due to factors such as the availability of channels increases with the number of handoff attempts. As the rate of handoff increases, handoff algorithms need to be enhanced so that the perceived QoS does not degrade and the cost to cellular infrastructure does not increase.

## 2. LITERATURE SURVEY

Existing work addresses the concept of Fixed Channel Allocation Scheme (FCA) where there are no separate channels allocated for handoffs. The available channels are shared by both new originating calls and handoff calls in first come first serve basis [8]. Here handoff request and new call request are dealt with equality. The cell doesn't consider the difference between Handoff request and new call request. It is intuitively clear that the termination of an ongoing call due to handoff failure is less desirable than the new call blocking. Hence the Quality of Service is not satisfied because the handoff blocking rate is as same as new call blocking rate. Many papers in the literature of related work address the categorization of the schemes based on guard channel concept [1]. The so called "Guard-channel" (GC) concept offers a means of improving the probability of a successful handoff by reserving a certain number of channels allocated exclusively for handoff requests. The remaining channels can be shared equally between handoff requests and new calls. Allocating Guard channels for Handoff improves the overall throughput which was discussed in our previous papers [10][11]. If the guard channel number is too big, the new call blocking rate will be high because several channels are set aside for handoff requests even when the traffic load is low. In this case, the resources are wasted by not serving either for handoff request or new call request. If the number is too small, the handoff blocking rate can't be guaranteed under high traffic load. So this scheme enhances the QoS by reducing the handoff blocking rate in a stable traffic load. While when the traffic load is changing periodically or dynamically due to big event or working rush hours, it is not flexible enough to get good QoS.

## 3. PROPOSED WORK

In this paper, the authors devise a scheme called Dynamic Guard Channel Allocation with Channel Borrowing Strategy (DGCA-CBS). In this scheme the channels for handoff requests are dynamically allocated based on the observation of certain past period of time in the network. Also, this scheme introduces a concept of channel borrowing in which the guard channels could be allocated to the new originating calls if they are unused. When a new originating call arises and if all the available channels are occupied it will check for the guard channels. If it is unused the new call will occupy the guard channel. The main aim is to utilize the available resources efficiently and also to balance the load in the network traffic. The following section presents the proposed scheme.

### 3.1. Dynamic Guard Channel Allocation with Channel Borrowing Strategy (DGCA-CBS)

A call being forced to terminate during the service is more annoying than a call being blocked at its start. Hence the handoff call blocking probability is much more stringent than new call blocking probability. Therefore, in the proposed scheme, priority is given to handoff requests by assigning $S_R$ channels exclusively for handoff calls among the S channels in a cell. The remaining $S_C$ $(=S-S_R)$ channels are shared by both originating calls and handoff requests. The selection of number of guard channels exclusively for handoff call is essentially important factor to get good Quality of Service. For different type of traffic load and mobility factor, different number of guard channels is needed to be allocated. The number of guard channels can't be fixed when the traffic load is changing with the time. Hence the guard channel allocation is dynamically changed by monitoring the traffic condition for certain time period. Also, the proposed scheme introduces the concept of channel borrowing strategy. In this strategy the exclusive guard channels are borrowed and allocated to new originating calls if the channel is unused. An originating call is blocked if channel is not available in the target cell. The channel allocation model is shown in figure 2.

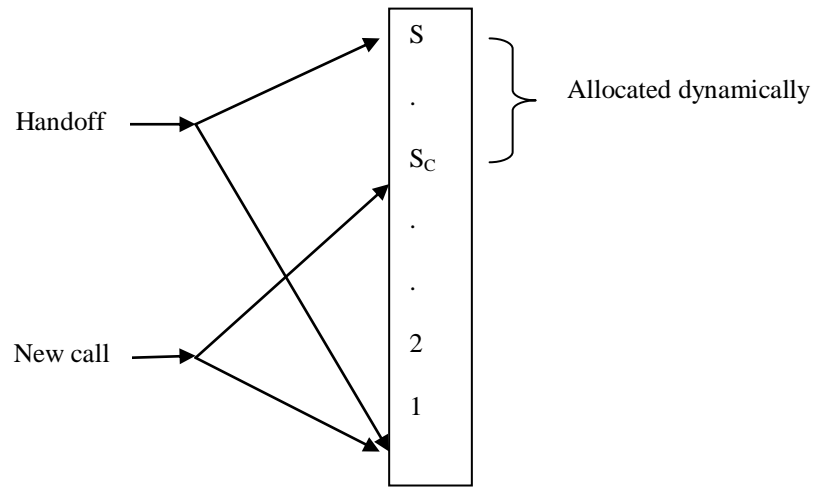

Figure 2. Dynamic Channel allocation model with priority for handoff calls.

In the proposed scheme, when a new call originates the BS will check for available channels in $S_C$. If channel is not free in $S_C$ it will check for non utilized channels in $S_R$. If channels are available in $S_R$ then BS borrows the channels and assigns it to the new originating calls. The BS monitors the traffic for certain time period and records it. Based on the network traffic the BS dynamically allocates the number of Guard channels exclusively for handoffs. Generally the number of handoff requests will be high in peak hours and is low during night and non peak hours. Hence based on the observed measurements the BS reallocates the guard channel. The entire operation is controlled by BS and MSC and hence the scheme is Network controlled handoff scheme [2].

As the number of Guard Channels allotted plays a vital role to the key performance, it is dynamically altered every specific time period say t. In this approach the number of guard channels which is to be allocated is determined through optimizing certain performance goal with service quality constraints. When a base station experiences high handoff blocking rate, the number of guard channels will be increased until the handoff blocking rate drops to below its threshold. When a base station does not get to use a significant portion of the guard channels over a period of time, the number of guard channels is gradually decreased until most of the guard channels are used frequently. By doing this, the handoff blocking rate is controlled to close to its threshold. The flow chart for the proposed scheme is depicted in figure 3.

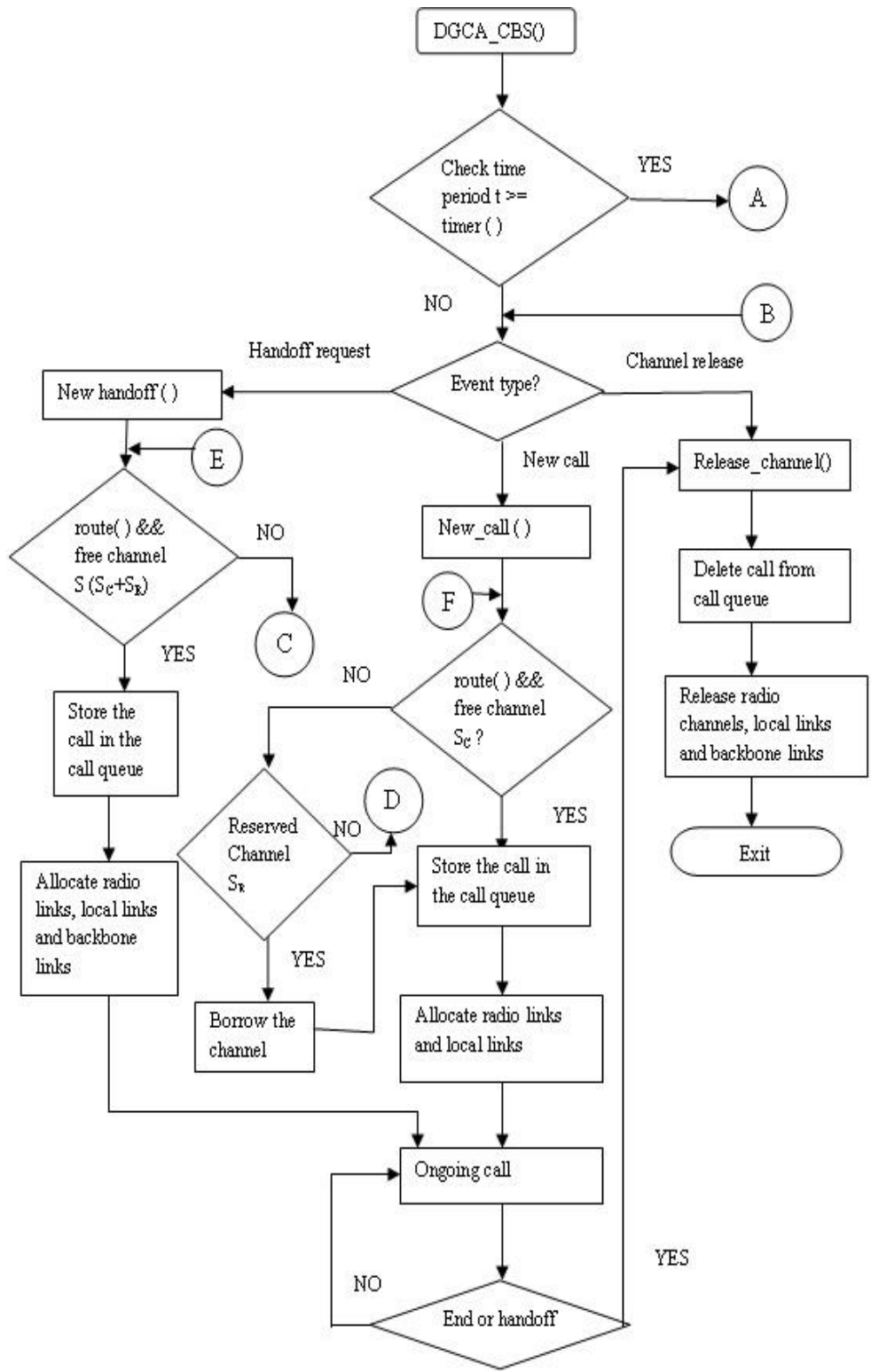

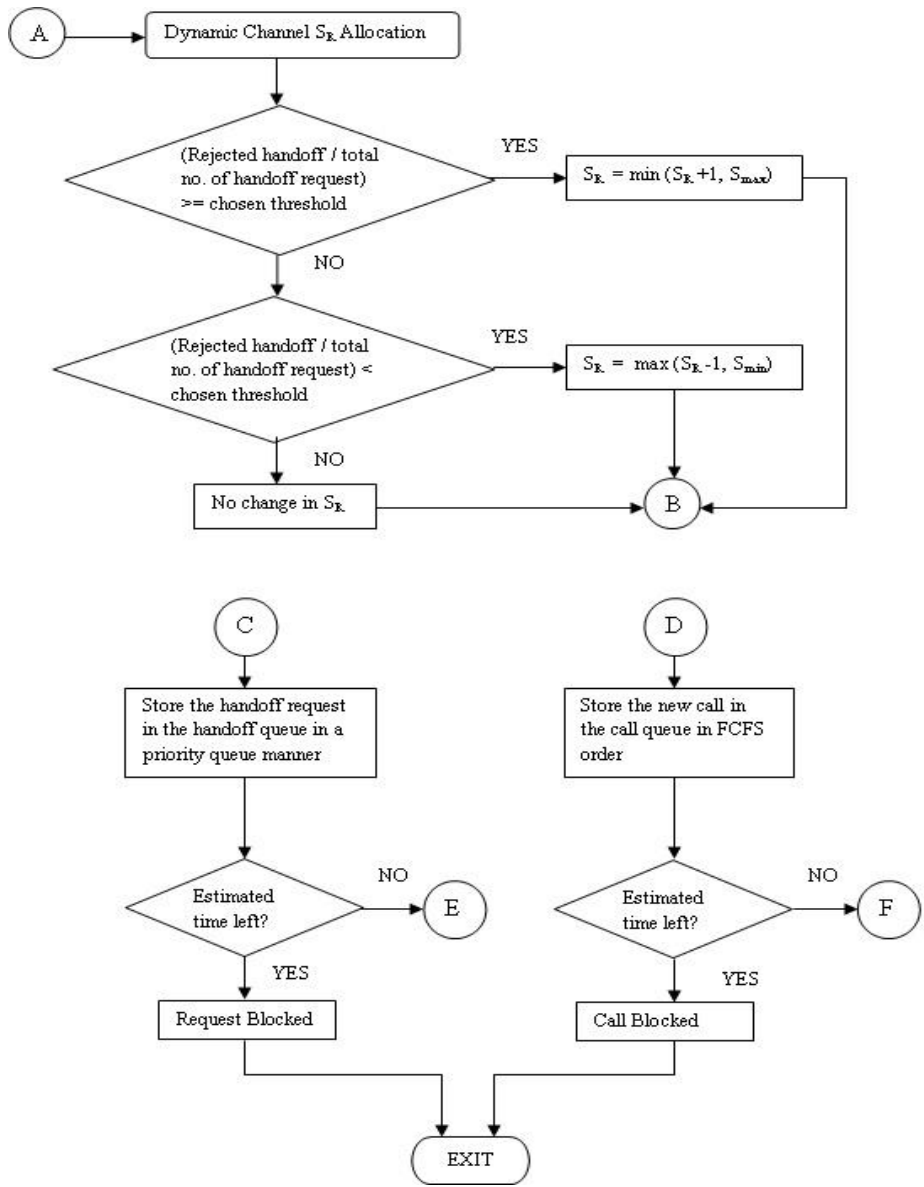

Figure 3. Flowchart represents DGCA-CBS

## 4. RESULTS AND DISCUSSION

The authors simulate the DGCA-CBS algorithm in six cells as a part of full network. The simulation program is implemented in Turbo C++, version 3.0, and run under MS DOS 6.2 environment. Object oriented approach is used to implement the real world environment. Results are directed to a text file and graphs for the same are obtained using MATLAB. The following graphs show the comparative study of the four schemes - Fixed Channel Allocation, Static Guard Channel Allocation, Dynamic Guard Channel Allocation without channel borrowing and Dynamic Guard Channel Allocation with Channel Borrowing Strategy.

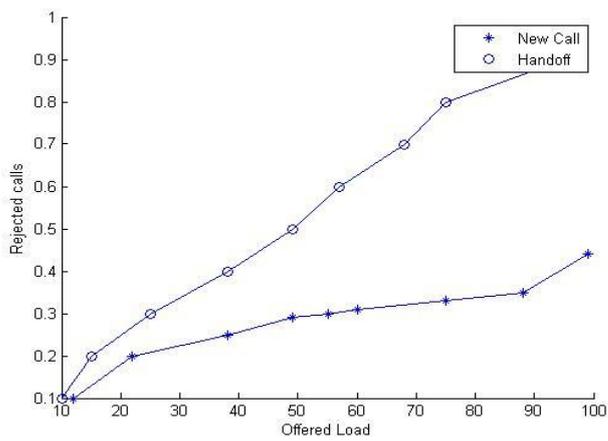

Figure 4. Fixed Channel Assignment scheme

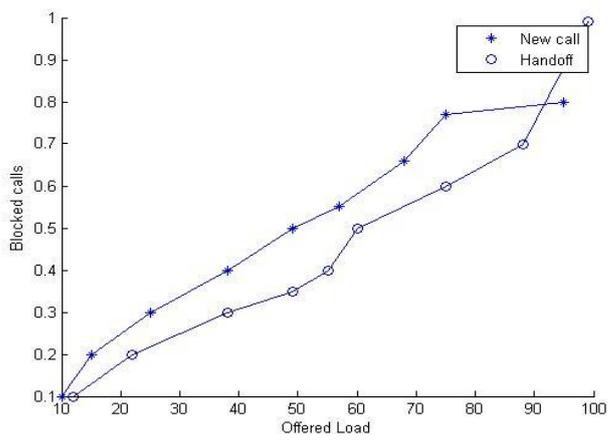

Figure 5. Static Guard Channel Allocation scheme

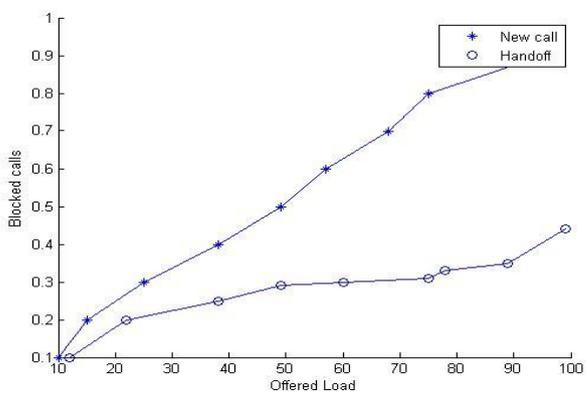

Figure 6. Dynamic Guard Channel Allocation without Channel Borrowing scheme

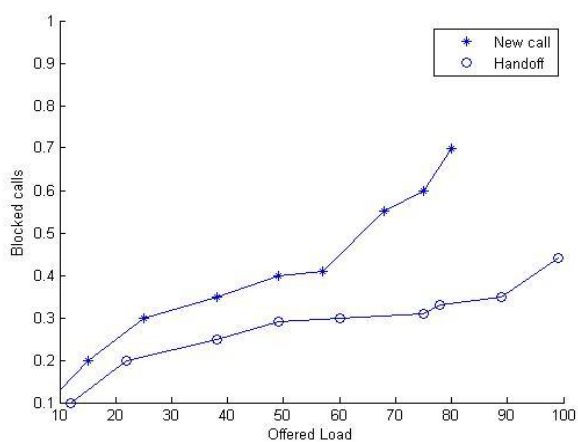

Figure 7. Dynamic Guard Channel Allocation with Channel Borrowing strategy (DGCA-CBS)

Figure 4 shows the simulated output of the Fixed Channel Assignment scheme where there is no separate guard channels allocated for handoff request. From the graph it is clear that handoff request rejection rate is high than the new call rejection rate. Hence the performance is not satisfactory as handoff requests should be given higher priority. Figure 5 shows the graph of Static Guard Channel Assignment scheme. This graph proves that the handoff blocking rate is reduced but the new call blocking rate is increased. Figure 6 shows the output of the Dynamic Guard channel allocation without channel borrowing scheme. From the graph it is apparant that the handoff blocking rate is considerably less than the new originating call blocking rate and the new call blocking rate is also better than graph in figure 5. Figure 7 shows the simulated output of the proposed scheme, Dynamic Guard Channel Allocation scheme with Channel Borrowing Strategy. From the graph it is evident that there is no difference for handoff calls when compared to the graph in figure 6, but for new call there is an improvement. The simulated output shows that there is a significant improvement in reducing the handoff request blocking rate and new call blocking rate in the proposed scheme DGCA-CBS, compared to the existing schemes. Hence there is a tradeoff.

## 5. CONCLUSION

In wireless mobile networks, as the cell size becomes smaller, handoffs occur more frequently. Careful design of call admission control scheme guarantees QoS to the mobile users. In this paper, the authors approached the channel allocation control scheme in a new way. A new Dynamic Guard Channel Allocation with Channel Borrowing Strategy (DGCA-CBS) is designed and implemented. The objective of this scheme is to utilize the available channel effectively. The main disadvantage of Dynamic Guard channel allocation scheme without channel borrowing strategy is new calls starving for channels while the guard channels remain unused. The concept of channel borrowing is incorporated in the proposed scheme which promises improvement in the new call blocking rate.

**Authors**

**Mrs.S.Alagu, M.Sc., M.Phil.,** currently a Research Scholar in Department of Computer Science and Engineering, Alagappa University, Karaikudi. She has a teaching experience of 10 years. She has published research papers in National and International Journals and Conferences. Her research area includes Wireless Mobile Networks.

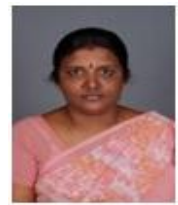

**Dr. T. Meyyappan M.Sc., M.Phil., M.B.A., Ph.D.,** currently, Associate Professor, Department of Computer Science and Engineering, Alagappa University, Karaikudi, TamilNadu. He has obtained his Ph.D. in Computer Science in January 2011 and published a number of research papers in National and International journals and conferences. He has been honored with Best Citizens of India Award 2012 by International Publishing House, New Delhi. He has developed Software packages for Examination, Admission Processing and official Website of Alagappa University. His research areas include Operational Research, Digital Image Processing, Fault Tolerant computing, Network security and Data Mining.

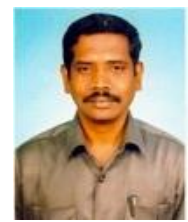